\documentclass[]{emulateapj}
\slugcomment{Astrophysical Journal Letters, in press}
\shorttitle{Spin-Orbit Alignment of HD\,17156b}
\shortauthors{Cochran et al.}

\begin{document}

\title{The Spin-Orbit Alignment of the HD\,17156
Transiting Eccentric Planetary System
\footnote{\rm Based
on observations obtained with the Hobby-Eberly Telescope, which is a joint
project of the University of Texas at Austin, the Pennsylvania State
University, Stanford University, Ludwig Maximilians Universit\"{a}t
M\"{u}nchen, and Georg August Universit\"{a}t G\"{o}ttingen.}
\footnote{\rm This paper includes data taken at The McDonald Observatory of
The University of Texas at Austin.}}

\author{William D. Cochran, Seth Redfield\altaffilmark{3}, Michael Endl
and Anita L. Cochran}
\affil{McDonald Observatory and Department of Astronomy,
The University of Texas at Austin, Austin TX 78712}
\email{wdc@astro.as.utexas.edu, sredfield@astro.as.utexas.edu,
mike@astro.as.utexas.edu, anita@barolo.as.utexas.edu}

\altaffiltext{3}{Hubble Fellow}

\begin{abstract}
We present high precision radial velocity observations of HD\,17156 during
a transit of its eccentric Jovian planet.
In these data, we detect the Rossiter-McLaughlin effect, which is an
apparent perturbation in the velocity of the star due to the progressive
occultation of part of the rotating stellar photosphere by the transiting
planet.   This system had previously been reported by \citet{NaSaOh08} to
exhibit a $\lambda = 62\arcdeg \pm 25\arcdeg$ misalignment of
the projected planetary orbital axis and the stellar rotation axis.  
We model our data, along with the \citeauthor{NaSaOh08} data, and obtain
$\lambda = 9.4\arcdeg \pm 9.3\arcdeg$ for the combined data set.   We thus
conclude that the planetary orbital axis is actually very well aligned with
the stellar rotation axis.
\end{abstract}


\section{Introduction}
Transiting extrasolar planets allow us to perform critical tests of models
of planetary system formation and evolution.
In our solar system, all of the planets orbit approximately, {\itshape but
not exactly}, in the solar equatorial plane.  \citet{BeGi05} find that the
angle between the plane of the ecliptic and the solar equator is
$7.155\pm0.002^{\circ}$.   The near co-planarity of the planetary
orbits in our solar system has influenced models of planetary system origin
from the time of \citet{Ka1755} and \citet{La1796} to essentially all
modern models \citep[e.g.][]{Li95,PoHuBo96,Bo00}.
Given the level of misalignment in our solar system,
and the observation of warps in debris disks around nearby stars such as
$\beta$~Pic \citep[e.g.][]{BuKrSt95,MoLaPa97,HeLiLa00}, it is obvious that
there are common processes which give rise to some small level of spin-orbit
misalignment in planetary systems.  The degree of misalignment in real
planetary systems must depend on the initial conditions (the initial
asymmetries of the collapsing cloud), the physics of disk formation
and evolution, and the physics of stellar mass and angular momentum
loss during the T~Tauri phase.  
Most transiting planets have periods of just a few days and orbits of low
eccentricity.   Significant exceptions are HD\,147506b (HAT-P-2b)
\citep{BaKoTo07}, with a 5.6 day orbital period and eccentricity of 0.52,
HD\,17156 \citep{FiVoMa07,BaAlLa07}, which has the longest orbital
period (21 days) and the largest eccentricity (0.67) of all of the
transiting exoplanets, and XO-3b, a massive (13.25 $M_J$) planet
which has an eccentricity of 0.26 in spite of it short orbital period of
3.192 days \citep{JKMCBu08}.
Scenarios for the formation of short-period planetary systems do not
necessarily deliver those planets to their present semi-major axes with low
eccentricity.   Type~II migration can result in planets with moderate
eccentricities \citep{SaGo04}, while dynamical interactions among newly
formed planets and planetesimals \citep{JuTr07} or the Kozai mechanism
\citep{HoToTr97,WuMu03} can result in large eccentricities and significant
inclination of planetary orbital planes from the plane of the stellar
equator (which is presumably close to the plane of the inner portion of the
proto-planetary disk).
While the shortest period transiting systems have probably been tidally
circularized, longer period systems can easily retain large
eccentricities over the main-sequence lifetime of the parent star.
A planet that has undergone significant gravitational scattering or
Kozai excitation would not necessarily retain a low inclination relative to
the stellar equator.
Thus, the report of a possible spin-orbit misalignment of the HD\,17156b
transiting planet by \citet{NaSaOh08} is extremely interesting,
and calls for further detailed investigation.
In this paper, we report our own spectroscopic observations of a transit
of HD\,17156 by its planet, using two telescopes at McDonald Observatory.

\section{Observations}\label{sec:obs}
We observed the transit of HD\,17156 by its hot-Jupiter companion on the
night of 25~December 2007~UT, using both the 2.7m Harlan J. Smith Telescope
(HJST) and the Hobby-Eberly Telescope (HET) at McDonald Observatory.
The HJST observations used the 2dcoud\'e spectrograph \citep{TuMQSn95}
in its ``F3'' mode, which gives a spectral resolving power of 
$R = \lambda/\delta\lambda = 60,000$.  This mode is referred to as ``cs23''.
A temperature stabilized I$_2$ gas absorption cell is used to impose the
velocity metric for precise radial velocity measurements of the stellar
spectrum.  Details of the 2.7m cs23 observing and data reduction
procedures are given by \citet{EnHaCo04,EnCoWi06}.  Velocity observations
were started before the expected beginning of the transit, and were continued
until after the expected end of the transit.  At total of 18~spectra, each
15~minutes in length, were obtained.  Table~\ref{tab:HJSvels}
gives the relative radial velocities for the HJST observations of HD\,17156. 

\begin{deluxetable}{rr@{.}lr}
\tablecolumns{4}
\tablecaption{McDonald Observatory 2.7m HJST Relative Velocities for HD\,17156
\label{tab:HJSvels}}
\tablehead{
\colhead{BJD}&
\multicolumn{2}{c}{Velocity}&
\colhead{$\sigma$} \\
\colhead{-2\,400\,000} &
\multicolumn{2}{c}{m\,s$^{-1}$} &
\colhead{m\,s$^{-1}$}}
\startdata
54459.604759 &-7898&41& 9.14 \\
54459.616367 &-7905&69& 9.79 \\
54459.627985 &-7936&46& 9.57 \\
54459.639520 &-7912&68&10.41 \\
54459.651060 &-7913&60& 8.61 \\
54459.663985 &-7921&94&12.61 \\
54459.675520 &-7916&51&11.56 \\
54459.687056 &-7936&95&10.14 \\
54459.698591 &-7955&00&11.86 \\
54459.710170 &-7960&05&12.44 \\
54459.721705 &-7985&06&10.98 \\
54459.733240 &-7964&48&12.75 \\
54459.744778 &-8003&38&10.36 \\
54459.757932 &-7977&30&11.57 \\
54459.769469 &-7979&22&11.89 \\
54459.781004 &-7988&41&12.40 \\
54459.792539 &-7990&68&10.82 \\
54459.804150 &-8010&27&10.81 \\
\enddata
\end{deluxetable}

HET observations were made on the same night (25~December 2007~UT) using
the High Resolution Spectrograph \citep{Tu98} in its $R = 60,000$ mode. 
Due to its fixed-zenith-distance design, we were only able to observe
HD\,17156 from shortly before the beginning of
the transit to just past past mid-transit.
The observations were planned to
obtain as many 600\,s exposures of HD\,17156 as possible during the 2.1 hour
track length.  The 13$^{th}$ target exposure on the HET was terminated
after 455\,s when the fiber-instrument-feed reached the end of track.
Details of the instrument configuration and the data reduction and analysis
procedures are given by \citet{CoEnMA04,CoEnWi07}.
Table~\ref{tab:HETvels} gives the relative radial velocities for
the HET observations of HD\,17156. 

\begin{deluxetable}{rr@{.}lr}
\tablecolumns{4}
\tablecaption{Hobby-Eberly Telescope HRS Relative Velocities for HD\,17156
\label{tab:HETvels}}
\tablehead{
\colhead{BJD}&
\multicolumn{2}{c}{Velocity}&
\colhead{$\sigma$} \\
\colhead{-2\,400\,000} &
\multicolumn{2}{c}{m\,s$^{-1}$} &
\colhead{m\,s$^{-1}$}}
\startdata
54459.605807 &  18&55 & 8.06 \\
54459.621927 &  10&35 & 8.01 \\
54459.629827 &   5&63 & 7.13 \\
54459.637718 &   0&74 & 8.11 \\
54459.645611 &  19&18 & 6.94 \\
54459.653504 &  11&45 & 6.32 \\
54459.661400 &   4&77 & 7.46 \\
54459.669293 &   4&65 & 7.34 \\
54459.677194 &  -4&16 & 7.34 \\
54459.685096 &  -7&11 & 6.32 \\
54459.692989 &  -9&91 & 7.68 \\
54459.700887 & -18&44 & 7.99 \\
54459.707944 & -35&93 & 9.25 \\
\enddata
\end{deluxetable}

For both the HJST and HET data, observation times and velocities have
been corrected to the solar system barycenter.   The uncertainty
$\sigma$ for each velocity in the table is an {\itshape internal} error
computed from the variance about the mean of the velocities from each of
the $\sim 2${\AA} small chunks into which the spectrum is divided
for the velocity computation.   Thus, it represents the relative
uncertainty of one velocity measurement with respect to the others for that
instrument, based on the quality and observing conditions of the spectrum. 
This uncertainty does not include other intrinsic stellar sources of
uncertainty, nor any unidentified sources of systematic errors.
The two different spectrographs have independent arbitrary
velocity zero points, and thus there is some constant offset velocity
(determined below and denoted as $\gamma$) between the data sets presented in
Tables~\ref{tab:HJSvels} and~~\ref{tab:HETvels}.

\section{Rossiter-McLaughlin Effect Model}\label{sec:RMModel}
A variety of different types of models have been used by others to analyze
observations of the Rossiter-McLaughlin (RM) \citep{Ro24,ML24}
effect for transiting planets.
\citet{QuEgMa00} divided a model stellar photosphere into a large number of 
cells, and then used a ``Gaussian shape cross-correlation model'' with a 
linear limb darkening law to compute the radial velocity anomaly.
\citet{OhTaSu05} developed analytic expressions for the apparent radial
velocity perturbation during a transit, in several different
approximations.
\citet{Gi06} developed another set of analytic expressions for the
RM effect which utilize a more generalized higher order
limb darkening expression.
A more elaborate technique was developed by \citet{WiNoHo05}, who
first computed an approximation to the disk-integrated stellar spectrum.
They then computed a Doppler shifted and intensity scaled spectrum of the
portion of the disk that would be blocked by the transiting planet and
subtracted this from their disk-integrated spectrum.  This spectrum was
then multiplied by their high-resolution iodine spectrum, and the result
was processed through their radial velocity code to compute model
velocities in the same manner as the observed data.

We analyzed our data using a model that is a hybrid of these methods.
We started by adopting the HD\,17156b system parameters from
\citet{NaSaOh08}.   We then computed the orbit of the planet around the
star, as we would view it from Earth.   This gave us the apparent offset of  
the planet from the center of the star as a function of time through the
transit.  We divided the stellar disk into a $400 \times 400$ grid of cells,
in the manner of \citet{QuEgMa00}, \citet{Sn04}, or \citet{WiNoHo05}.
For each photospheric cell, we computed a specific intensity using
the non-linear four-parameter limb darkening law of \citet{Cl00}.  Each
photospheric cell is also assigned a radial velocity due to both the stellar 
orbital motion and the stellar rotation, with the stellar
$v \sin i$ as a model parameter.
For each time step during the transit, from first contact to fourth
contact, we compute which stellar photospheric cells are blocked by
the transiting planet.  
We then integrate
the unblocked Doppler-shifted and intensity weighted stellar photospheric
cells to compute both the RM radial velocity perturbation
during the transit and the transit photometric lightcurve.

We fully recognize the limitations and approximations inherent in this
modeling procedure.
First, there is no {\itshape a priori} reason to assume that limb-darkening
should follow any particular law.   Also, the limb darkening in photospheric
absorption lines is quite different from the limb darkening in the
continuum.   While this appears to be taken into account by \citet{Cl00},
the limb darkening parameterization is based on model atmospheres rather
than on real stars.   More importantly, in our model we compute the
specific-intensity weighted apparent Doppler shift of the visible
portion of the photosphere.   On the other hand, our spectra record
transit-perturbed stellar absorption line profile shapes from which
we measure an apparent Doppler shift using a computer code that assumes an
unperturbed line-profile shape.  In future improvements to our
RM model, we will attempt to improve several of these limitations.

\section{Data Analysis}\label{sec:analysis}

We used the model described in Sec.~\ref{sec:RMModel} to analyze
simultaneously the data sets from the HJST cs23, the HET HRS, and
the observations published by \citet{NaSaOh08}, which we will refer
to as the ``OAO/HIDES'' data set.
Since each data set has its own independent velocity zero-point,
we allowed the systemic velocity of each
data set to be an independent free parameter in the analysis.
The values of the fixed parameters for the analysis are given in
Table~\ref{tab:fixedparams}.  The planetary orbital elements were taken from 
\citet{IrChNu08}.   These elements are essentially indistinguishable from
the single-planet fit of \citet{ShWeOr08}.
We note that the conclusions of this work depend on the particular values
of these parameters we adopt.  If any of these parameters turn out to be in
significant error, those errors will propagate through this analysis.

\begin{deluxetable}{lr@{.}ll}
\tablecolumns{4}
\tablecaption{Assumed System Parameters \label{tab:fixedparams}}
\tablehead{\colhead{Parameter}&\multicolumn{2}{c}{Value}&\colhead{Source}}
\startdata
$M_\star$ ($M_\sun$)         & 1&2     & \citet{FiVoMa07} \\
$R_\star$ ($R_\sun$)         & 1&47    & \citet{FiVoMa07} \\
\makebox[3em][l]{Claret} a1  & 0&5346  & \citet{Cl00} \\
\makebox[3em][l]{}       a2  & 0&1041  & \citet{Cl00} \\
\makebox[3em][l]{}       a3  & 0&4189  & \citet{Cl00} \\
\makebox[3em][l]{}       a4  &-0&2584  & \citet{Cl00} \\
$R_p$ ($R_{Jup}$)            & 1&01    & \citet{IrChNu08} \\
$P$ (days)                   &21&21691 & \citet{IrChNu08} \\
$T_p$ (HJD)             &2453738&605   & \citet{IrChNu08} \\
$K$ (m\,s$^{-1}$)           &273&8     & \citet{IrChNu08} \\
$e$                          & 0&670   & \citet{IrChNu08} \\
$\omega$ (degrees)         & 121&3     & \citet{IrChNu08} \\
\enddata
\end{deluxetable}

In modeling the data, we allowed the orbital plane inclination~$i$,
the projected angle~$\lambda$ between the planetary orbital axis and the
stellar rotation axis, the projected stellar rotational velocity $v \sin i$,
as well as the systemic velocity of each separate data set to be free
parameters.  We then minimized the chi-squared of the model fit to the data.

For the combined data sets, we obtained $\lambda = 9.4\arcdeg \pm
9.3\arcdeg$ at $v \sin i = 6.3 \pm 1.1$\,km\,s$^{-1}$ and
$i = 85.9\arcdeg \pm 0.4\arcdeg$.
This indicates that the planet is orbiting near the stellar
equatorial plane, to within the uncertainty of our determination.
Our derived $v \sin i$ is somewhat larger than the value of
4.7\,m\,s$^{-1}$ found by \citet{NaSaOh08} and 
the 2.6\,km\,s$^{-1}$ given by \citet{FiVoMa07}.
Our inclination is within one $\sigma$ of the \citeauthor{NaSaOh08}
value of $85.65\arcdeg$.
However, our value for $\lambda$ from the combined data differs
significantly from that of \citet{NaSaOh08},
who found $\lambda = 62\arcdeg \pm 25\arcdeg$.  

\begin{deluxetable*}{lr@{$\pm$}lr@{$\pm$}lr@{$\pm$}lccrr}
\tablecolumns{3}
\tablecaption{Rossiter-McLaughlin Model Results \label{tab:results}}
\tablehead{\colhead{Data Set}&
\multicolumn{2}{c}{$\lambda$}&
\multicolumn{2}{c}{$i$}&
\multicolumn{2}{c}{$v \sin i$}&
\colhead{$\chi^2$}&
\colhead{degrees of}&
\colhead{$\gamma_c$}&
\colhead{$\gamma_i$}\\
\colhead{}&
\multicolumn{2}{c}{(degrees)}&
\multicolumn{2}{c}{(degrees)}&
\multicolumn{2}{c}{(km\,s$^{-1}$)}&
\colhead{}&
\colhead{freedom}&
\colhead{m\,s$^{-1}$}&
\colhead{m\,s$^{-1}$}
}
\startdata
Combined      &   9.4 &  9.3 & 85.9 & 0.4 & 6.3 & 1.1 & 40.53 & 49 & \\
HJST cs23     &   4.5 & 15.6 & 86.1 & 0.6 & 7.1 & 2.1 & 13.19 & 14 & -7853.0 & -7854.0 \\
HET HRS       & -32.4 & 25.2 & 86.3 & 0.5 & 4.8 & 1.5 &  3.71 &  9 &    73.3 &    65.6 \\
OAO/HIDES     &  59.4 & 19.7 & 86.2 & 0.8 & 8.6 & 1.9 & 14.25 & 21 &   142.1 &   146.5 \\
\enddata
\end{deluxetable*}

In order to understand the reason for this difference,
we then modeled each of the three data sets separately.
The derived $\lambda$ and $i$ for the combined fit, as well as for
each individual data set are given in Tab.~\ref{tab:results}.
We also give $\gamma_c$, the systemic velocity for each data set
in the combined fit, as well as $\gamma_i$, the systemic velocity for
each individual data set fit.
From the OAO/HIDES data set alone, we computed 
$\lambda = 59.4\arcdeg \pm 19.7\arcdeg$ at $i = 86.2\arcdeg \pm
0.8\arcdeg$.  If we fix the inclination at the $i = 85.65\arcdeg$  
value of \citet{NaSaOh08}, we get $\lambda = 61.4\arcdeg \pm 28.5\arcdeg$.
The surprisingly excellent agreement of all of these value for the
OAO/HIDES data set thus validates the modeling process of
\citeauthor{NaSaOh08}.
We note that the results from the HJST data
alone are in very good agreement with the combined results. 
Due to the design limitations of the HET, 
the HET/HRS data only covered the first half of the transit.
Thus, the code could trade-off $\lambda$ vs. the systemic velocity for the
data set, and derived a slightly lower $\chi^2$ with 
$\lambda = -32.4\arcdeg \pm 25.2\arcdeg$
for the HET data alone.

\begin{figure}
\plotone{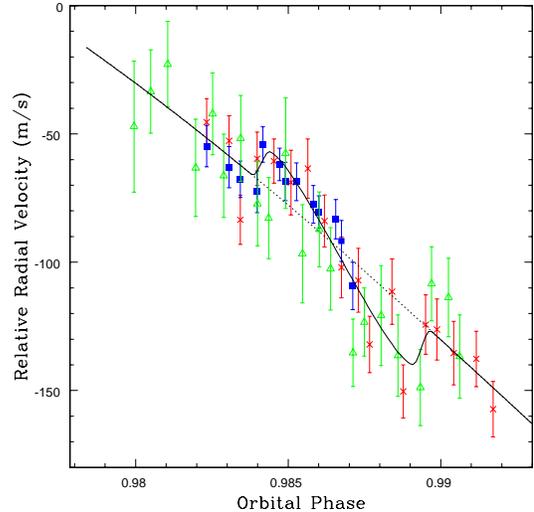}
\caption{The fit of our Rossiter-McLaughlin effect model (solid black line)
to the observational data sets.  The HJST cs23 data are shown as the
(red) Xs, the HET HRS data are (blue) solid squares, and the OAO/HIDES data are
(green) open triangles.  The dashed black line is the orbital velocity of
the star in the absence of any Rossiter-McLaughlin velocity perturbation.
\label{fig:rvfit}}
\end{figure}

The model fit to the data is shown in Fig.~\ref{fig:rvfit}.
We also computed a model with no RM effect
by setting $v \sin i = 0$.   This removes the stellar rotation, and thus
there is no apparent Doppler shift of the portion of the stellar
photosphere eclipsed by the planet.
The $v \sin i = 0$ model gave $\chi^2 = 71.56$ ($\chi^2_r = 1.43$)
as opposed to $\chi^2 = 40.53$ ($\chi^2_r = 0.83$)
for the best fitting RM model for the combined data set. 
Thus, we conclude that the RM effect was indeed
convincingly detected in spite of the apparent noise in the data.  
Examining the $\chi^2$ of each individual data set in the full model
and the $v \sin i = 0$ model showed that even in the noisiest case
of the OAO/HIDES data, the RM effect reduced
the total $\chi^2$ by 13.1.

Due to the symmetric signature of the RM effect
in the $\lambda \approx 0$ case, it is critically important to sample the
entire transit.  This is evidenced by the fit to the HET/HRS data alone,
in which despite the high quality of the data, only the first half of the
transit was observable, and therefore erroneous parameters are derived.
In order to successfully calibrate the baseline
radial velocity variation, observations well before and after the transit
are required.  If the radial velocity can be well calibrated by
observations before and after transit, the influence of uncertainty in
the zero-point velocity offset ($\gamma$) will be negligible on the
RM effect parameters.

\section{Discussion}\label{sec:conclusions}
We have reanalyzed the data of \citet{NaSaOh08} along with our own
independent observations of the spectroscopic transit of the eccentric
exoplanet HD\,17156b, and we find that $\lambda = 9.4\arcdeg \pm 9.3\arcdeg$.   
We conclude that this exoplanetary system is similar to almost all
of the other short-period transiting exoplanetary systems studied using the
RM effect so far, in that it shows that the projected 
planetary orbital axis appears to be aligned with the projected stellar
rotation axis, to within our measurement precision.
The only remaining notable exception is XO-3b, which was reported by
\citet{HeBoPo08} to show a very significant misalignment of
$\lambda = 70\arcdeg \pm 15\arcdeg$.

The HD\,17156b system is extremely interesting because it has a very high
eccentricity ($e = 0.67$) for such a short orbital period of 21.2\,days.
Most other transiting systems have much shorter orbital periods and
eccentricities near zero.   Many of the interesting ways of getting a
planet into this type of orbit might well result in the possibility of an
orbital plane significantly inclined to the stellar equatorial plane.
Dynamical interactions with another nearby planet could result in ejection
of the other body and a large orbital inclination of the surviving body.

The presence of a third planet in the system, as suggested by
\citet{ShWeOr08}, would have some effect on the orbital elements
of HD\,17156b.  However, as \citet{ShWeOr08} discussed, the primary effects
are a small oscillation of the eccentricity and a secular advance of the
argument of periastron.   As long as the two planets are approximately
coplanar, there would be no induced change in the orbital inclination.
Thus, it appears that the dynamical process that placed HD\,17156b 
into a short-period highly eccentric orbit did not significantly affect the
orbital inclination of the system with respect to the stellar equatorial
plane.    

\acknowledgments  \vspace{-15pt}
We are extremely grateful for receiving Director's Discretionary time on the
HET on very short notice in order to obtain the data
presented here.
S.R. would like to acknowledge support provided by NASA through Hubble
Fellowship grant HST-HF-01190.01 awarded by STScI,
which is operated by AURA Inc., for NASA, under contract NAS 5-26555.
This material is based on work supported by NASA under Grant NNG05G107G
issued through the TPF Foundation Science Program, and under
Cooperative Agreement
NNA06CA98A issued through the {\itshape Kepler} Program.
The Hobby-Eberly Telescope (HET) is a joint project of the University of
Texas at Austin, the Pennsylvania State University, Stanford University,
 Ludwig Maximilians Universit\"{a}t M\"{u}nchen, and
Georg August Universit\"{a}t G\"{o}ttingen.  The HET is named in honor
of its principal benefactors, William P. Hobby and Robert E. Eberly.

{\it Facilities:} \facility{HET}, \facility{McD:2.7m}.

\end{document}